\documentclass[prc,aps,final,twocolumn,nofootinbib]{revtex4-1}
\usepackage{graphicx}
\usepackage{mathrsfs}
\usepackage{bm}
\usepackage{colordvi}

\newcommand{\bel}[1]{\begin{equation}\label{#1}}

\usepackage{epsfig,psfrag}

\def\ramaSM{\vadjust{\vbox to 0pt{\vss \hbox to \hsize
{\hskip\hsize \quad $\Leftarrow$\quad {\it SM}\hss}\vskip3.5pt}}}
\def\ramaSL{\vadjust{\vbox to 0pt{\vss \hbox to \hsize
{\hskip\hsize \quad $\Leftarrow$\quad {\it SL}\hss}\vskip3.5pt}}}
\def\ramaSR{\vadjust{\vbox to 0pt{\vss \hbox to \hsize
{\hskip\hsize \quad $\Leftarrow$\quad {\it SR}\hss}\vskip3.5pt}}}
\def\rama{\vadjust{\vbox to 0pt{\vss \hbox to \hsize
{\hskip\hsize \quad $\Leftarrow$\quad
{$\Longleftarrow$}\hss}\vskip3.5pt}}}

\def\be{\begin{equation}}
\def\ee{\end{equation}}
\def\bea{\begin{eqnarray}}
\def\eea{\end{eqnarray}}
\def\l{\label}

\def\hahat{\hat{H}}

\def\hahat0{\hat{H}_0}

\def\cos{\hbox{cos}}

\def\exp{\hbox{exp}}

\def\siml{\hbox{\kern.1em \lower.6ex \hbox{$\sim$} \kern-1.12em
          \raise.6ex \hbox{$<$} \kern.1em }}
\def\simg{\hbox{\kern.1em \lower.6ex \hbox{$\sim$} \kern-1.12em
          \raise.6ex \hbox{$>$} \kern.1em }}

\def\siml{\hbox{\kern.1em \lower.6ex \hbox{$\sim$} \kern-1.12em
 \raise.6ex \hbox{$<$} \kern.1em}}
\def\simg{\hbox{\kern.1em \lower.6ex \hbox{$\sim$} \kern-1.12em
 \raise.6ex \hbox{$>$} \kern.1em}}

\newcommand{\beqar}{\begin{eqnarray}}
\newcommand{\eeqar}[1]{\label{#1} \end{eqnarray}}

\begin{document}

\title{Two-neutron transfer 
  reactions and
  quantum-chaos measure of 
nuclear spectra}
\bigskip
\author{A.I. Levon \email{Email: levon@kinr.kiev.ua}}
\author{A.G. Magner
\email{Email: magner@kinr.kiev.ua}}
\affiliation{
  Institute for Nuclear Research, 03680 Kyiv, Ukraine}
\bigskip
\begin{abstract}

    A new statistical interpretation of the nuclear collective states is
suggested and applied
recently in rare earths and 
actinide nuclei by the two-neutron transfer reactions in terms of 
the nearest neighbor-spacing distributions (NNSDs). Experimental NNSDs
were obtained by using the complete and pure sequences of the collective
states through an unfolding procedure.
The two-neutron transfer reactions 
 allow to obtain such a sequence of the collective states that 
meets the requirements for a statistical analysis.
Their theoretical analysis is 
based on the 
linear approximation of 
a repulsion level density 
within the Wigner-Dyson 
theory. This approximation 
is successful to evaluate separately the Wigner chaos 
and Poisson order contributions. We found an intermediate behavior of NNSDs 
between the Wigner and Poisson limits. NNSDs 
turn out to be shifted from a 
chaos to order with increasing the length of spectra and the angular 
momentum of collective states. Perspectives for the statistical analysis 
of the symmetry breaking of states 
with the fixed projection of angular 
momenta $K$ are discussed.

\end{abstract}

\maketitle

\noindent

\section{Introduction}

For last two decades the analysis of the energy spectra of nuclei, atoms and 
other many-body quantum system becomes very attractive \cite{1,2,3,4,5}. 
The quantum 
chaos measure plays a central role for understanding the
universal properties of energy spectra for such a quantum 
system. As these properties belong to 
the whole spectrum of a given many-body system, and they are too complicate 
for using simple models based on the model Hamiltonian \cite{6,7}, the 
statistical methods can be applied successfully (see, e.g., the review 
\cite{5}). 
A constructive idea for improving statistics is to compile sequences of 
states having the same quantum numbers in several nuclei, - e.g., angular 
momentum and parity. For this purpose, one can use averaged distances
between nuclear levels for the scale transformation of energy states.

Different statistical methods have been proposed to obtain information on 
the chaoticity versus regularity in quantum spectra of a nuclear many-body 
system \cite{1,2,3,4,5}, see also the well known work by 
Bohigas, Giannoni and Schmit 
\cite{8}. The short-range fluctuation properties in experimental spectra can be 
analyzed in terms of the nearest-neighbor spacing distribution (NNSD) 
statistics. The uncorrelated sequence of energy levels, originated by a 
regular dynamics, is described by the Poisson distribution. In the case of a 
completely chaotic dynamics, the energy intervals between levels follow 
mainly the Wigner (Gaussian orthogonal-ensemble, GOE)
distribution. An intermediate degree of chaos in energy spectra is usually 
obtained through
a comparison of the experimental NNSDs with well known distributions 
\cite{9,10,11,12,13}
based on the fundamental works \cite{8,13,14,15}. This comparison 
is carried out 
\cite{16,17,18,19,20} by using the least square-fit technique. 
The estimated values of 
parameters of these distributions shed 
light on the statistical 
situation with considered spectra. Berry and Robnik \cite{14} derived the NNSD 
starting from the microscopic semiclassical expression for the level density 
through the Hamiltonian for a classical system.
The Brody NNSD \cite{10} is based 
on the expression for the level repulsion density that interpolates between 
the Poisson and the Wigner distribution by only one parameter. 

\begin{figure*}
\vskip1mm
\includegraphics[width=14.0cm]{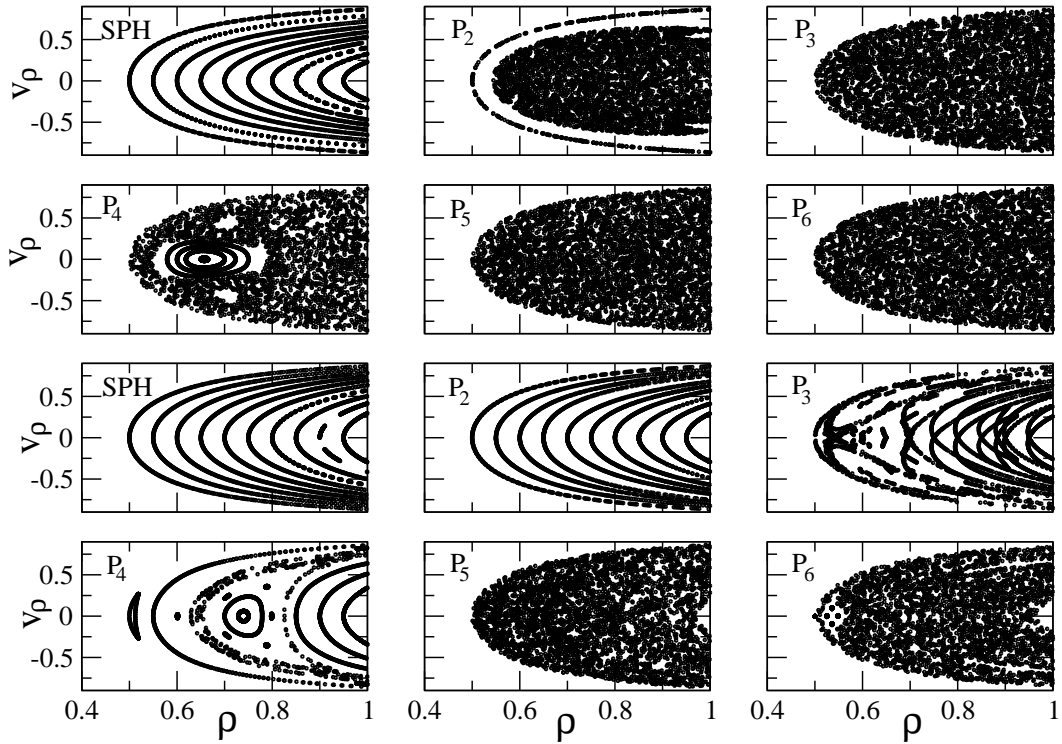}

\vspace{1.5cm}
\vskip-3mm\caption{{\small
 Poincare sections $\rho ,v_{\rho }$ ($v_{\rho }=p_{\rho }/m$ is 
the perpendicular velocity and m the mass of particle) for spheroidal (SPH) 
cavity and 5 axially-symmetric shapes $r=R[1+\alpha P_{L} (\cos \,\vartheta 
)]$ with indices $L=2,3,4,5$ of the Legendre polynomials $P_{L} (\cos 
\,\vartheta )$ in the spherical coordinates $r,\vartheta ,\varphi $ by 
accounting for all projections of the angular momenta $K$; two last lines: 
$\alpha =0.005$; two upper lines: $\alpha =0.4$.
}}
\label{fig1}
\end{figure*}

For a quantitative measure of the degree of chaoticity of the many-body 
dynamics, the statistical probability distribution $p(s)$ as 
function of spacings 
$s$ between the nearest neighboring levels can be derived within the general 
Wigner-Dyson (WD) approach based on the level repulsion density 
$g(s)$ (the units 
will be specified later) \cite{1,2,13,15}. This approach can be applied in the 
random matrix theory, see for instance \cite{3,15}, and also, for systems with 
a definite Hamiltonian \cite{1,2}.
In any case, the order in such systems is 
approximately associated with the Poisson dependence of $p(s)$ on the 
spacing $s$ 
variable, that is obviously related to a constant $g(s),$ independent of $s$. A 
chaoticity can be referred, mainly, to the Wigner distribution for 
$g(s)\propto s$. 

For a further study of the order-chaos properties of nuclear systems, it 
might be worth to apply a simple analytical approximation to the WD NNSD 
$p(s),$ keeping the link 
to a level repulsion density $g(s) $
\cite{1,2,15}. For analysis of 
the statistical properties in terms of the Poisson and Wigner distributions, 
one can use the linear WD (LWD) approximation to the level repulsion density 
$g(s)$ \cite{21,22}. It is the two-parameter approach; in contrast, 
e.g., to the 
one-parameter Brody approach \cite{10}.
However, the LWD approximation, as based on a 
smooth analytical (linear) function $g(s)$ of $s$, 
can be 
derived
properly within the 
WD theory (see Refs.~\cite{1,2,22}). 
Moreover, it gives a more precised information on 
the separate Poisson order-like and Wigner chaos-like contributions. Within 
a linear level-repulsion density $g(s),$ the NNSD $p(s)$ was reduced to one 
parameter and, at the same time, the same quantitative information of their 
order and chaos contributions was keeping in Ref.~\cite{23}.
One of the most 
attractive questions is a change of the statistical structure of NNSDs  
by the symmetry breaking due to the fixed projection of
the angular momentum of 
collective states to the symmetry axis. For the case of the single-particle 
(s.p.) states, see for instance \cite{21}. 

In the present 
review we discuss the application of NNSDs 
\cite{1,2,5,9,11,12,13,14,15,21,22,23} to analyze the experimental data 
\cite{24,25,26,27,28,29}. 
This article is organized as the following. In Sec.~II we 
consider the general semiclassical characteristics of the
classical and quantum chaos in 
many-body systems \cite{30,31,32}.  
Experiments \cite{24,25,26,27,28,29} 
based on the two-neutron transfer reactions are analyzed in Sec.~III.
An unfolding scale-transformation procedure for calculations of the 
experimental NNSDs 
of nuclear states in heavy complex nuclei is discussed in Sec.~IVA, see 
also Refs.~\cite{16,17,18}. 
In Sec.~IVB, a short review of the theoretical Wigner-Dyson 
approaches for simple NNSD calculations 
\cite{1,2,9,13,21,22,23} is presented. They 
are used for the statistical analysis of the obtained experimentally 
\cite{24,25,26,27,28,29}
collective-excitation spectra in Sec.~V, in contrast to
that of the s.p.
spectra \cite{5,20}. The paper is ended by a summary.

\section{Regular and chaotic systems}

\begin{figure*}
\vskip1mm
\includegraphics[width=14.0cm]{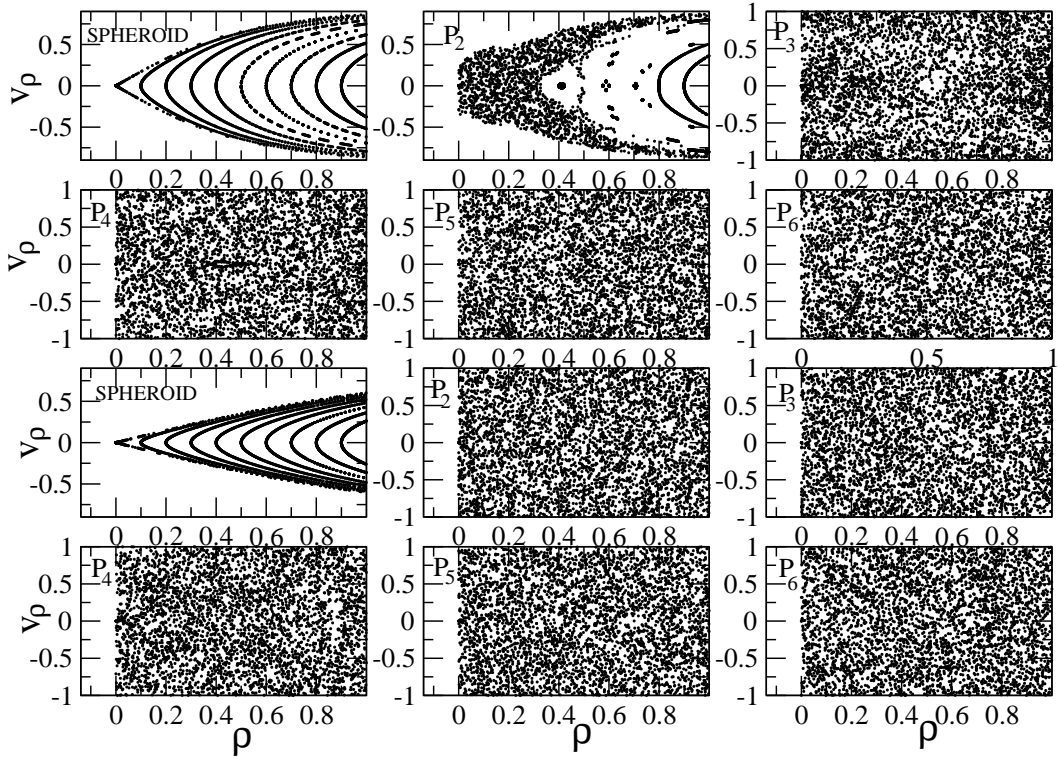}

\vspace{1.2cm}
\vskip-3mm\caption{{\small
 The same as in Fig.~\ref{fig1} 
but for the fixed projection of the  
angular momentum $K=0 $; two upper lines: $\alpha =0.005$; 
two last lines: 
$\alpha =0.4$.
}}
\label{fig2}
\end{figure*}

A particle motion in the Hamilton classical mechanics is called regular 
(integrable) if a small perturbation of initial conditions induces a small 
deflection of the classical trajectory from the initial phase-space point. 
In contrast to this, such a motion is chaotic if a behavior and time 
evolution of the complex classical many-body system depend essentially from 
the initial conditions, even with small variations of the surrounded medium. 
In this case, even a small perturbation of the initial phase-space point
produces a significant influence on the trajectory solutions of the 
Hamiltonian system and lead, thus, to an exponential increasing 
deflections of the classical trajectory. This can be described 
quantitatively in terms of the so called classical Lyapunov exponent. 
Another transparent way of the classical chaos-order analysis is to 
calculate the Poincare sections (see Figs.~\ref{fig1} and \ref{fig2} from
Ref.~\cite{32}) as perturbations of the initial phase-space 
point in the direction perpendicular to a given periodic orbit (PO)
and to study, 
then, their evolution after several periods of the particle motion 
along this referred non-perturbative PO
\cite{31,32}.

It might be seemed from the first view that there is no such chaos-order 
transitions in the quantum mechanics based on the deterministic Hamiltonian. 
According to the Heisenberg uncertainty principle in the Feynman 
path-integral formulation \cite{30,31,33,34}, one can say only about the 
probability of finding a trajectory of particle motion to calculate a 
Green's function propagator of the wave function. Integrating over all 
of intermediate phase-space points along such formal trajectories, 
one can evaluate the probability to find a particle at a given phase-space 
point as function of time.

\begin{figure}
\vskip1mm
\includegraphics[width=8.0cm]{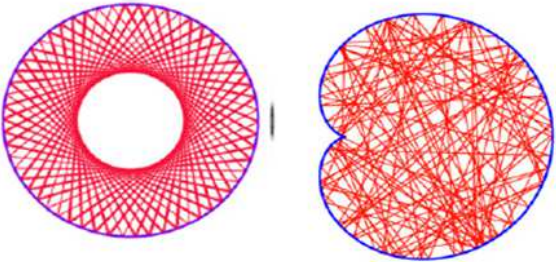}

\vskip-3mm\caption{{\small
 Demonstration of the particle motion in spherical and heart classical 
billiards.
}}
\label{fig3}
\end{figure}
\begin{figure*}
\includegraphics[width=14.0cm]{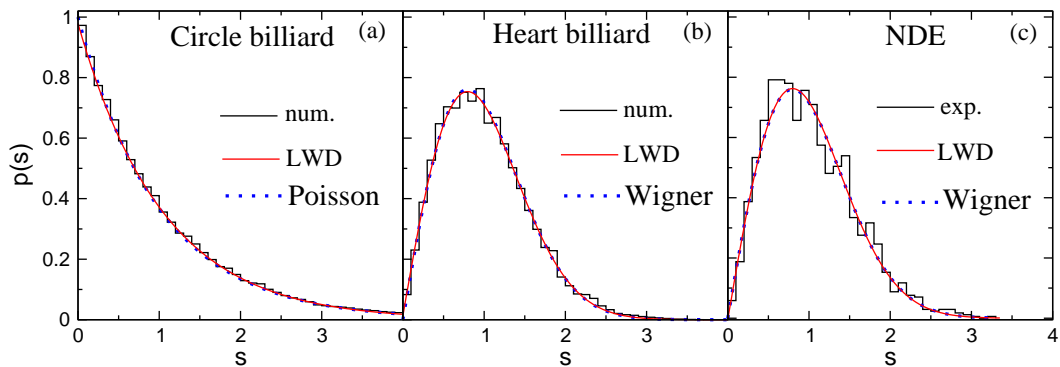}

\vspace{1.2cm}
\caption{{\small
     NNSDs $p(s)$ as functions of a dimensionless spacing variable $s$ for
     (a) Poisson- and (b) Wigner-like numerical calculations and (c)
     Wigner-like results (see text) by staircase lines.
     One-parametric LWD (\ref{11}) are
    shown by solid lines. Dots present the Poisson (a) and Wigner (b,c)
    (\ref{1}).
}}
\label{fig4}
\end{figure*}
\begin{figure*}
\includegraphics[width=14.0cm]{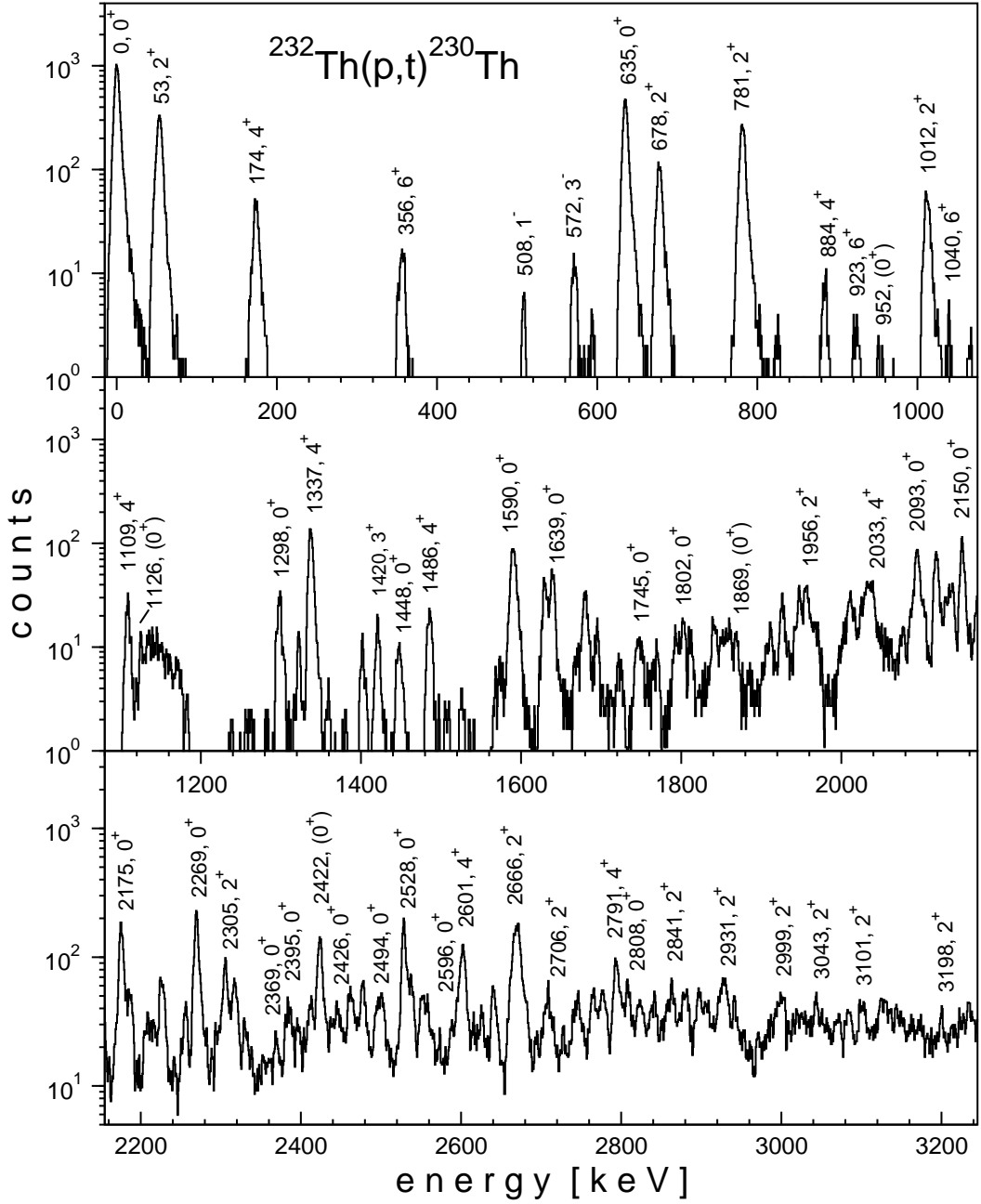}

\vspace{1.5cm}
\caption{{\small
 Spectrum for the $^{234}U(p,t)^{232}U$ reaction (in logarithmic 
scale) for a detection angle of 5$^{\circ}$. Most of the 
levels are labeled with their excitation energy in keV.
}}
\label{fig5}
\end{figure*}
\begin{figure*}
\includegraphics[width=14.0cm]{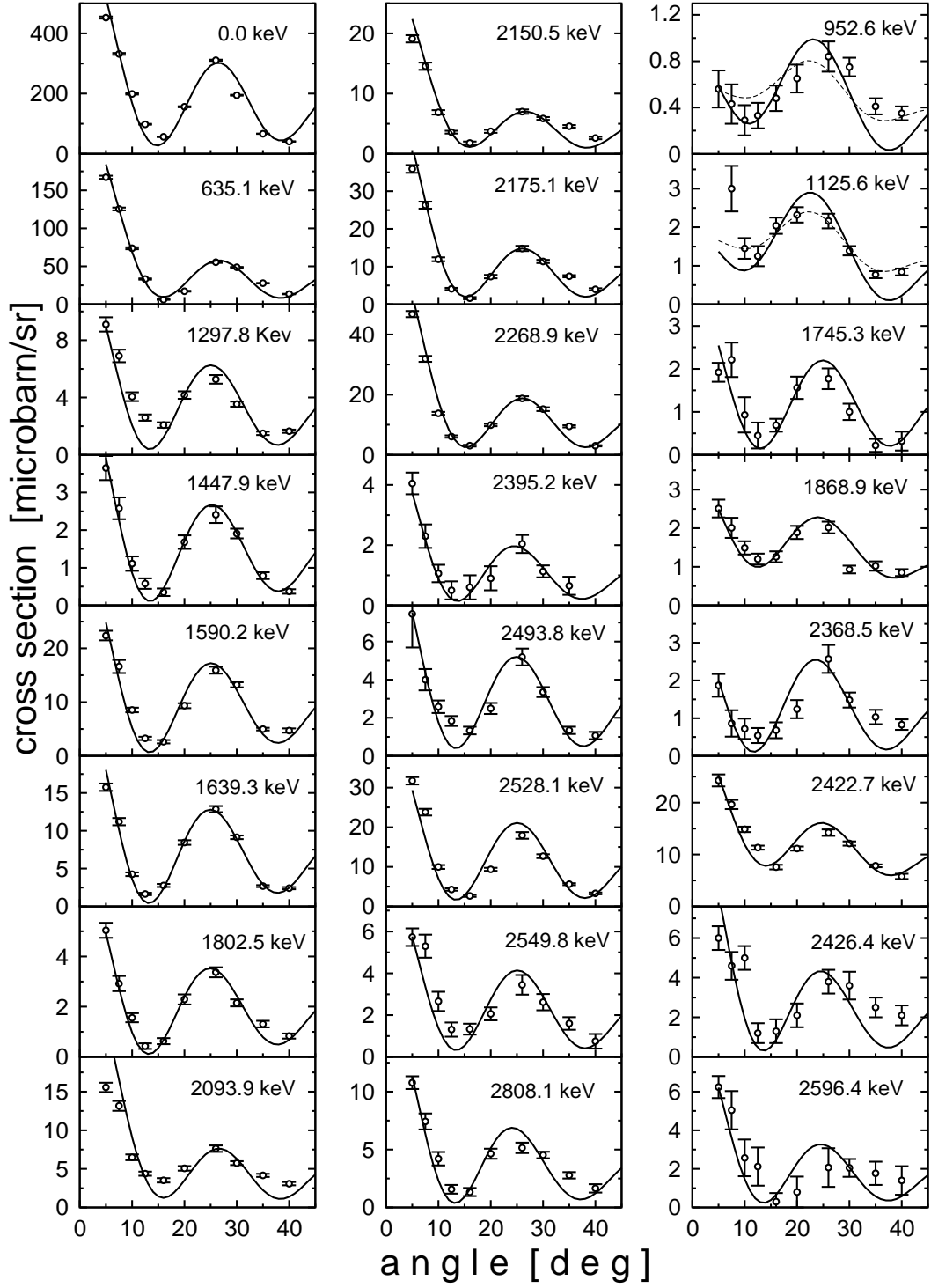}

\vspace{1.5cm}
\caption{{\small
 Angular distributions of assigned 0$^{+}$ states in $^{230}$Th and 
their fit with CHUCK3 one-step calculations. Dashed lines show fits
for 1$^-$ states as possible alternative assignments.
}}
\label{fig6}
\end{figure*}
\begin{figure*}
\includegraphics[width=14.0cm]{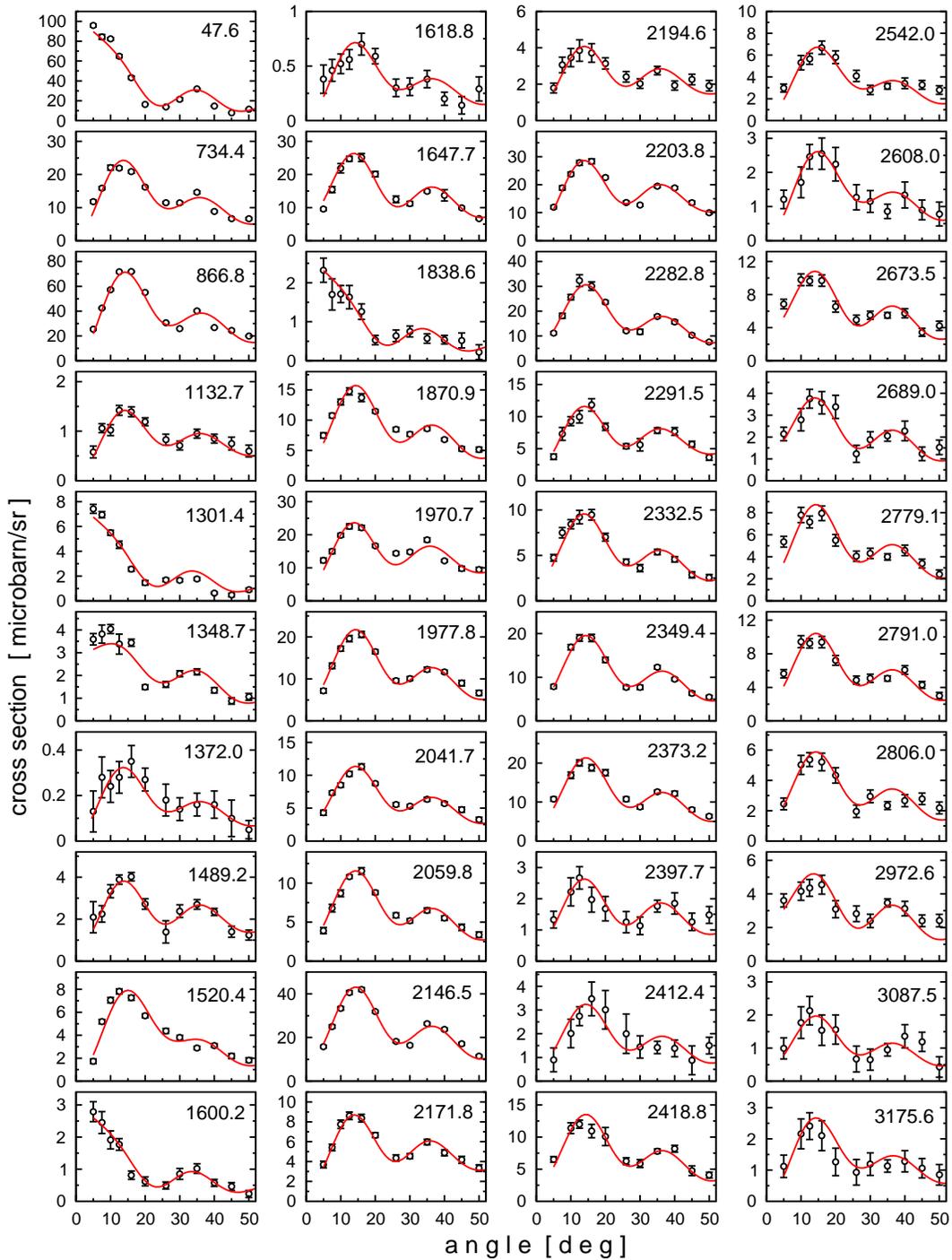}

\vspace{1.0cm}
\caption{{\small
 Angular distributions of assigned 2$^{+}$ states in $^{232}$U and 
their fit with CHUCK3 calculations (labels yield excitation energies in 
keV).
}}
\label{fig7}
\end{figure*}
However, there is an obvious relationship between symmetries of the 
Hamiltonian potential
used in classical and quantum mechanics. The number of 
independent
parameters for a given
particle action constant, except for the energy $E$, is the same in both these 
formulations, i.e., the same symmetry parameter $\mathcal{D}$ 
which determines a degree 
of chaos \cite{33,35,36,37}. A bridge between these two different classical and 
quantum approaches is the semiclassical 
PO theory 
\cite{30,31,33,34,35,36,37,38,39,40} where the Lyapunov exponent 
$\lambda $ appears in 
terms of the eigenvalues $\exp (\lambda )$ and $\exp (-\lambda )$ of the 
stability matrix $\rm{M}_{\rm{st}}$ ($\mbox{det}\rm{M}_{\rm{st}}=1$). 
This matrix is a propagator for 
the time evolution of perturbations , e.g., in two dimensions $\delta \xi 
=\left(\delta \rho , \delta p_{\rho } \right)$ in 
the perpendicular direction $\rho $ to a given referred 
PO after 
several periods of particle motion from the initial $\delta \xi^\prime$
to the final  $\delta \xi^{\prime\prime}$ perturbation,
$\delta \xi^{\prime\prime}=$\rm{M}$_{\rm{st}}$ 
$\delta \xi^{\prime}$ \cite{32}. 
A regular motion 
is described by restricted values of $\lambda(t)$ for finite perturbations 
$\delta \xi(t)$ while a chaotic motion is related to its increasing 
values $\delta \xi(t)$ in time. Similarly, the quantum-classical 
correspondence can be described in terms of the Poincare sections shown in
Figs.~\ref{fig1} and \ref{fig2}~- the 
phase space points $\left(\rho^{\prime\prime} ,p_{\rho }^{\prime\prime}\right)$ after 
many periods of particle motion along the reference PO
starting from the initial point
  $\left(\rho^{\prime},p_{\rho }^{\prime}\right)$
\cite{31,32}. 
In a completely integrable system, all classical trajectories are  
POs, e.g., in the harmonic oscillator with rational ratios of frequencies 
\cite{31,33,34,40}, where the symmetry parameter $ \cal{D}$ is maximal, 
equal to 2n-2 
for n degrees of freedom. This leads to a completely degenerate level 
density as sum over PO
families
\cite{40}. 
If the energy $ E$ is one 
single-valued integral of motion ($\mathcal{D}=0$) for a
completely non-integrable 
Hamiltonian, one has a full chaotic behavior of the classical trajectories 
and, relatively, a
discrete sum over isolated
POs in the semiclassical density of 
quantum states \cite{30,31,33,34,35,36,37}. Fig.~\ref{fig1} 
shows transparently the increasing of 
chaos for the transitions from the integrable spheroidal cavity to the 
chaotic Hamiltonian systems with increasing Legendre polynomial index $L $ and 
deformation parameter $\alpha $. Thus, there is no contradictions between 
the classical and the quantum chaos description for the same deterministic 
Hamiltonian because of a bridge by the semiclassical 
PO theory.

Other famous phenomena are the symmetry restoring and symmetry breaking 
in a particle system described by the Hamiltonian with potential depending 
on a parameter like the 
deformation parameter $\alpha $ \cite{31,33,34,35,36,37,39}. Fig.~\ref{fig2} 
shows a shift to the 
chaoticity with the fixed s.p. angular momentum projection $K$ as compared to 
all of mixed projections in Fig.~\ref{fig1}. 
This shift is enhanced much with 
increasing the deformation of the system $\alpha $.

In the case of the symmetry restoration, one has to mention the bifurcations 
\cite{36,37,39,41} as catastrophe values of the deformation parameter $\alpha $ 
in Fig.~\ref{fig1}. 
This shift is growing 
with increasing the system deformation 
$\alpha $ , where one solution of the classical Hamiltonian equations 
is transformed to two solutions with a local increase of the symmetry 
parameter $\cal{D} $ and, therefore, a shift to a regular behavior. 
In the opposite 
- symmetry breaking - case, one of the parameters $\cal{D}$ is fixed, say, the 
projection of the angular momentum $K$ in a band of the collective states with 
a given total angular momentum $J$ and parity $\pi $. This restriction of the 
phase space should lead to the increasing of chaos with respect to the order 
behavior \cite{21} due to a lost of the Hamiltonian symmetry. As seen 
from comparison between the Poincare sections of Figs.~\ref{fig1} 
and ~\ref{fig2}  
with fixing $K$ (Fig.~\ref{fig2})  
for a given polynomial $P_{L\, }$ and deformation $\alpha $, the 
Poincare sections become obviously more chaotic than with accounting for all 
the angular momentum projections $K$  (Fig.~\ref{fig1}),
i.e., the chaoticity
measure increases with fixed 
$K$. Note also that, as expected, the chaoticity is enhanced with 
non-integrability and complexity of the shapes.

Thus, evidence for quantum chaos can be obtained by using the correspondence 
principle with respect to the classical chaos. In this relation, one can 
transparently consider the two simplest billiard systems: spherical and 
cordial billiards where the regular and chaotic behavior of classical 
trajectories takes place respectively, see Fig.~\ref{fig3}. 
 Let us deal 
now with the corresponding quantum billiards as a system of 
independent particles moving in a cavity potential. 
The probability to find such a potential is 
determined by the wave function squared as a solution of the Shr\"{o}dinger 
equation. Solving the corresponding eigenvalue problem, one can find the 
quantum spectrum and study its statistical properties. The Orsay group 
accumulated sequences of many (of the order of 1000) eigenvalues which 
belong to the eigenfunctions of the same symmetry (e.g., the same angular 
momentum $J$ and parity $\pi $). Numerical calculations, as well as 
experiments, provide the finite energy-level sequences of the whole spectrum 
for a quantum system. The question is how to obtain the relevant 
statistical properties of these sequences by comparing them with the 
appropriate statistical theory. Fig. \ref{fig4} shows (very) good
agreement between the 
numerical calculations of the
NNSDs $p(s)$ within the Wigner-Dyson theory \cite{22,23} for the
circle (a) and cordial (b)
billiards as functions of the spacing variable $s$ (in dimensionless 
units of the local energy level distances)  as compared to 
the Poisson regular and Wigner chaotic distributions, respectively, 
\begin{equation}\label{1}
p(s)=\exp (-s),
\qquad
p(s)=\frac{\pi s}{2}\,\exp \left( {-\frac{\pi \,s^{2}}{4}} \right).
\end{equation}
In Fig.~\ref{fig4} (c)
  the nuclear data ensemble (NDE),
  which includes 1726 neutron and proton resonance
  energies, is found also in good agreement with the Wigner
  distribution of Eq.~(\ref{1}).
NNSDs will be considered below in more details after a presentation of
the relevant experimental data on the two-neutron (p,t) reactions 
  like shown in Fig.~\ref{fig5}.

\section{Two-neutron transfer reactions}

To perform a statistical analysis of energy spacings, one needs the complete 
and pure level sequences. The completeness means absence of missing and 
incorrectly identified energy levels. For nuclear physics, this requirement 
is to use the levels of identical symmetries: The level sequences have to be 
used with identical angular momenta $J$ and parity $\pi $. Additional quantum 
numbers 
can be considered in some problems, for instance, the
isospin $T $ or the 
angular momentum projection $K$ to the symmetry axis. For a 
statistical evidence, the level sequences should be enough long. These 
conditions are satisfied in the spectra obtained by using the reactions with 
a two neutron transfer. As an example, see the proton-triton 
reaction spectrum for the target $^{234}U$ at the angle 5$^{\circ}$ 
(Fig.~\ref{fig5}).

The energy spectra were measured for 10 angles in the range of 5 - 40 
degrees and, thus, the angular distributions for each excitation 
level were obtained (see Figs.~\ref{fig6} and \ref{fig7}). 
To get information on the angular 
momenta $J$ and parity $\pi $ for the observed levels, the angular 
distributions were analyzed by using the coupled channel method through
the program 
CHUCK3 based on the distorted wave Born approximation (DWBA).
Multi-step calculations include a two-neutron transfer and 
 excitations in the same nucleus (up to 8 ways). The initial aim of 
such experiments was investigations of the nature of multiple 0$^+$ 
excitations. Spectra of 2$^+$, 4$^+$ and 6$^+$ states were obtained as a 
secondary information which was turned out to be useful in the present 
statistical analysis.

\begin{figure}
\includegraphics[width=7.5cm]{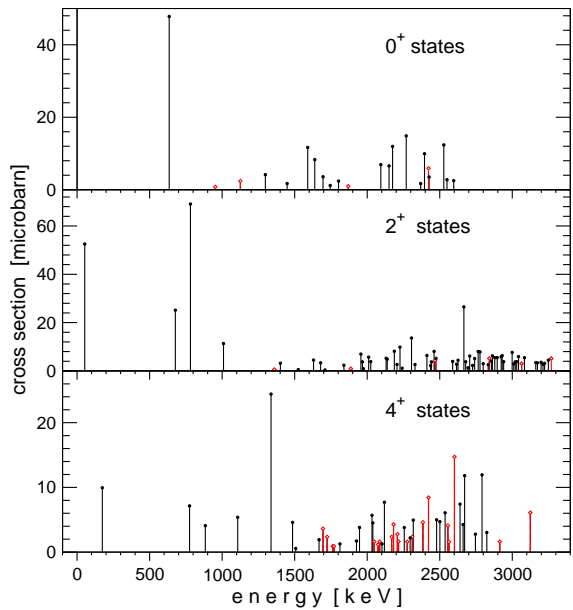}

\vspace{0.2cm}
\caption{{\small
 Experimental distribution of the (p,t) strength integrated in the 
angle region 
0$^{\circ} $-- 45$^{\circ} $ for 0$^{+}$, 2$^{+\, }$ and 4$^{+\, }$ states
in $^{230}$Th. 
The levels identified reliably and  tentatively are indicated by filled
circles and 
open diamonds, respectively.
}}
\label{fig8}
\end{figure}

Figs.~\ref{fig6} and \ref{fig7} 
demonstrate the quality of
 experimental results and of their 
analysis. The final results of such study are shown in Fig.~\ref{fig8} 
for the 
$^{230}$Th nucleus. The energies, spins, parities and cross sections for 
each level are determined. They are 
combined for each given angular momentum.

In the framework of the problem under consideration, it is important to have 
information on the nature of states excited in the two-neutron transfer 
reaction. It was shown that, at least, the 0$^+$, 2$^+$, 4$^+$ and 6$^+$ states 
are collective. Some of evidences of the collective nature of these states 
are given below.

\begin{figure*}
  \includegraphics[width=6.0cm]{Fig9a_expincrQPM.eps}
~~
  \includegraphics[width=6.4cm]{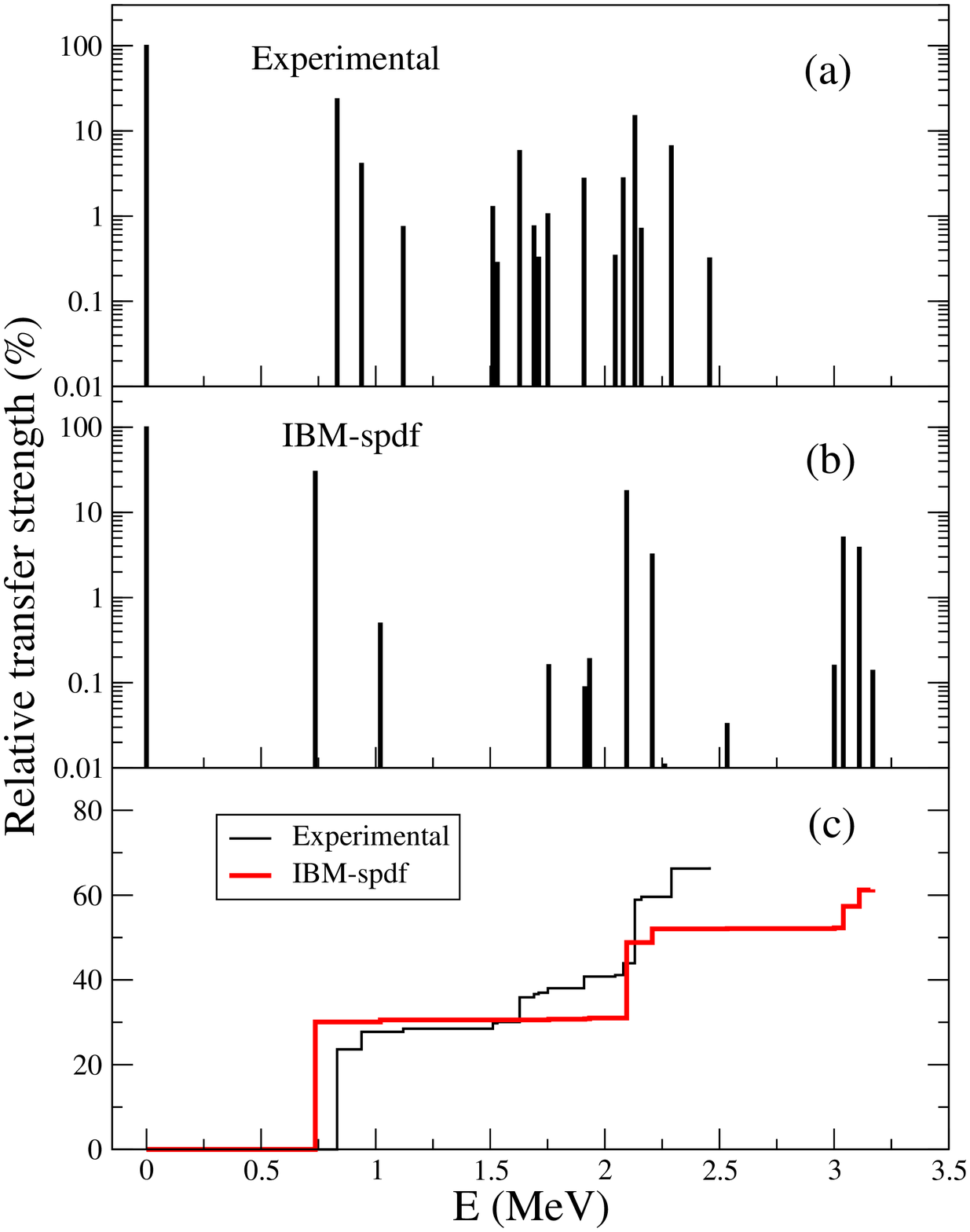}

\vspace{0.5cm}
\caption{{\small
    Experimental increments of the (p,t) strength in $^{\, 230}$Th (left)
and (p,t) strengths for 
0$^{+\, }$ states in $^{228}$Th (right) 
are compared with the QPM and IBM (a, b) calculations, relatively;
experimental versus 
computed cumulative 
sums of the (p,t) strength 
are given on right (c).
}}
\label{fig9}
\end{figure*}

Theoretical calculations of the energies, cross sections, and structure of 
the states excited in the two-neutron transfer were carried out within the 
framework of a quasiparticle-phonon model (QPM \cite{6}) and 
the interacting boson model (IBM [7]).
Both models give the absolute cross sections 
which are close to experimental ones. Fig.~\ref{fig9} 
demonstrates 
good agreement of 
the experiment and calculations in frame of the QPM on left and IBM on 
right. Cumulative pictures of the experimental and theoretical spectroscopic 
factors, are rather similar. As to the nature of these states, in all the 
low-lying states, quadrupole phonons are dominant with a relatively modest 
role of the octupole phonons. The contribution of the latter increases with 
the growing excitation energy.

\vspace{2.5cm}
\begin{figure*}
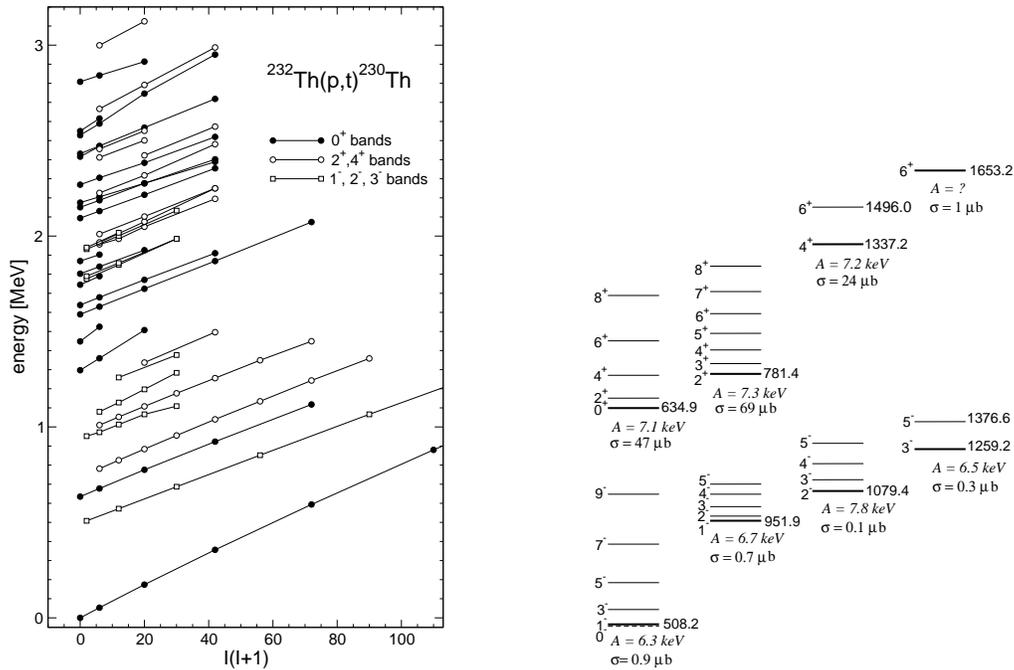

\hspace{-0.5cm}\includegraphics[width=5.8cm]{Fig10a_collbands.eps}
~
\hspace{1.5cm}\includegraphics[width=5.8cm]{Fig10b_collbands.eps}

\vspace{1.0cm}
\caption{{\small
    Collective bands based on 
    0$^{+}$, 2$^{+}$, 4$^{+}$, 1$^{-}$, 
2$^{-} $ and 3$^{-} $ excited states in $^{230}$Th as assigned from the DWBA fit 
of 
angular distributions (Figs.~\ref{fig6} and 
\ref{fig7}) 
from the (p,t) 
reaction (left), and assumed multiplets of states of the octupole one-phonon 
(bottom) and 
two-phonon (top) 
energies in keV, 
associated with the 
collective bands (right);
$\sigma $ is the
cross section
in microbarns. 
}}
\label{fig10}
\end{figure*}
%

\vspace{-2.5cm}
The phenomenological IBM turned up to be successful in explaining the 
experimental results of the two-neutron transfer reactions within the 
\textit{spdf}-IBM version using the Extended Consistent Q-formalism 
\cite{42}.  This model 
gives spectra of 0$^{+}$, 2$^{+}$, and 4$^{+}$ states which are close to the 
experiment, and their excitation cross sections, as well as the ratios of 
reduced transition probabilities $B(E1) / B(E2)$.

\vspace{1.5cm}
\begin{figure}
\includegraphics[width=7.0cm]{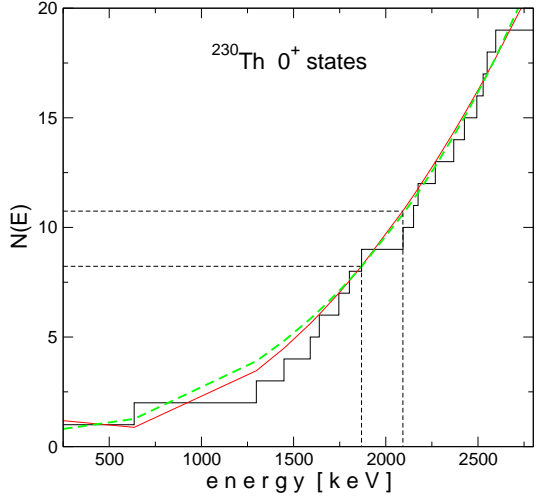}

\vspace{0.2cm}
\caption{{\small
    Histogram of the cumulative states' numbers 
    $N(E)$ and 
    their  fitting by two polynomials 
      (\ref{4}) (green dashed) and (\ref{5})
    (red solid)
    for the 0$^{+}$ 
    energy spacings in $^{230}$Th.
}}
\label{fig11}
\end{figure}

\vspace{-0.5cm}
\begin{figure*}
\includegraphics[width=14.0cm]{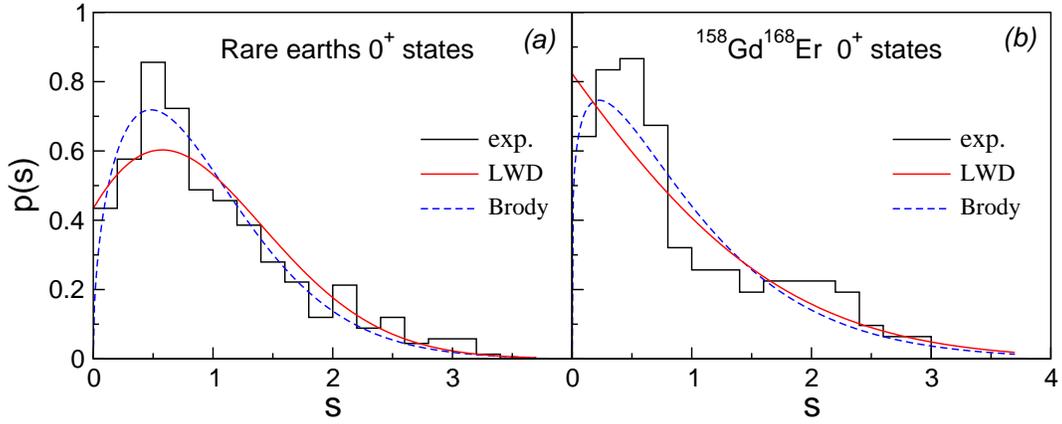}

\vspace{0.2cm}
\caption{{\small
        NNSDs $p(s)$ (staircase 
    line) as functions of the 
    spacing $s$  for 0$^{+}$ 
collective states and 
fits by the LWD (\ref{11}) 
and the Brody (\ref{7}) approach
for 
energies $E<3$MeV in many rare nuclei  
(a) and for 
energies $E < 4.2$MeV in$^{158}$Gd and $^{168}$Er 
nuclei (b).
}}
\label{fig12}
\end{figure*}
\begin{figure*}
\includegraphics[width=14.0cm]{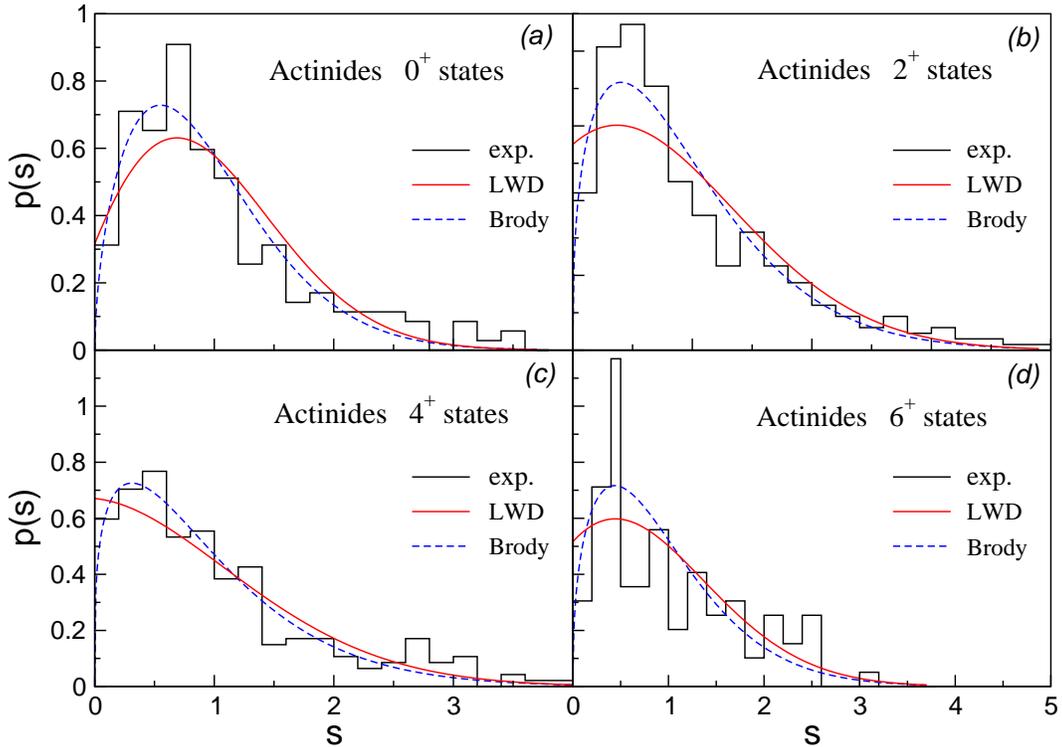}

\vspace{0.5cm}
\caption{{\small
    The same as in Fig.~\ref{fig11} 
 but for different states in 
 actinide 
 nuclei: 0$^{+}$,~2$^{+},~$4$^{+}$ and 6$^{+}$.
}}
\label{fig13}
\end{figure*}
%

\vspace{-1.0cm}
For example, Fig.~\ref{fig9} 
(right) shows the experimental and calculated spectra of 0$^{+}$ states
{\bf (a,b)} and 
the experimental increment of the (p,t) strength {\bf (c)} in comparison with the 
theoretical ones. In the structure of a part of these states, in addition 
to the \textit{sd}-bosons, an important role plays also \textit{pf}-bosons.
 By other words, 
octupole excitations are essential. Thus, the collective nature of states 
excited in the $(p,t)$- reaction is confirmed in this model, too.

Another evidence of the collective nature of states excited in the 
two-neutron transfer reaction are rotational bands, that can be built from 
the identified states. After the assignment of spins to all excited states, 
the sequences of states, which can be distinguished, show the 
characteristics of a rotational band structure.

An identification of the 
states associated with rotational bands was made on the following 
conditions:

 \vspace{0.2cm}
 i) The angular distribution for a state as the band member candidate is
 assigned by the DWBA calculations for the spin, that can be necessary
 to put into the band;

\vspace{0.2cm}
ii) The transfer cross section in the (p,t) reaction to the states in
the potential band has to be 
decreased with the increasing spin;

\vspace{0.2cm} 
iii) The energies of states in the band can be fitted approximately 
by the expression for a 
rotational band $E =  E_{0}  + \mathcal{A}J(J+1)$ with a 
$E_{0}$ constant, and a small and smooth variation of the inertial 
parameter $\mathcal{A}$.

\vspace{0.2cm}

Collective bands identified in such a way are shown in Fig.~\ref{fig10}.
Under the above criteria (i)-(iii), the 
procedure can be 
justified for some sequences. 
They are already known 
from gamma-spectroscopy to belong to the rotational bands. The straight 
lines in Fig.~\ref{fig10} strengthen the arguments for these assignments.

Finally, multiplets of states are identified in the actinide nuclei which 
can be treated as quadruplets of one- and two-phonon octupole states. Since 
the octupole degree of freedom plays an important role in this mass region, 
such a result was expected though the identification of the two-phonon 
octupole quadruplet was obtained for the first time. Both quadruplets are 
shown in Fig.~\ref{fig10} (right).

The levels excited in the 
two-neutron transfer reaction and 
identified in the way above described 
are included into the analysis of 623 
states, see Ref.~\cite{22}. 

\section{Theoretical approaches to NNSDs}

\subsection{Unfolding procedure}

 To compare properly 
the statistical properties of different sequences to each other, one should 
convert any set of the energy levels into a set of the normalized spacing, 
that can be done through the so-called unfolding procedure \cite{8,22}.  
In this procedure the original sequence of level energies $E_{i}$ is 
transformed to a new dimensionless sequence $\varepsilon_{i} $ 
($i=1,2,...$ numerate the levels) as mapping
\begin{equation}\label{2}
\varepsilon_{i} =\tilde{{N}}(E_{i} ),
\end{equation}
where $\tilde{N}(E)$ is a smooth part of the cumulative level density,
\begin{equation}\label{3}
N(E)=\int\limits_0^E {d{E}'\,dN({E}')/d{E}'} \quad ,
\end{equation}
with the level density $dN(E)/dE$. As shown in Fig.~\ref{fig11}, 
the 
cumulative density $N(E)$ is the staircase function that counts the number 
of states with energies less or equal to $E$. Usually, a polynomial of not 
large order is used to fit $N(E) $. In Fig.~\ref{fig11} we tested the two
polynomials,
\begin{equation}\label{4}
  \tilde{N}(E)=a^{}_{0} +a^{}_{1} E+a^{}_{2} E^{2},~
\end{equation}
and
\begin{equation}\label{5}
\tilde{N}(E)=a^{}_{0} +a^{}_{1} E^{2}+a^{}_{2} E^{4},
\end{equation}
and found small differences for the corresponding fitting.
In such a way, 
the spectra will be analyzed in terms of 
the spacings between the unfolded energy levels (\ref{2}),
\begin{equation}\label{6}
s_{i} = \varepsilon_{i+1\, }-\varepsilon_{i\, } . 
\end{equation}
The NNSD is, then, the distribution of a probability $p(s)$ 
to find the number of 
unfolded levels $\Delta N$ in the interval $\Delta s$.

\vspace{-0.7cm}
\subsection{Analytical NNSD approximations} 

NNSDs~~ $p(s)$~~are defined as the probability distribution, i.e., 
the probability to find 
a level between $s$ and $s+ds$. As it is the neighbor 
levels, this NNSD is, first 
of all, a quantitative measure of chaos and regularity for close correlations. 
The spectral fluctuations are not described by nor Wigner
(GOE)
Poisson limits (\ref{1}), 
i.e., the system is both not pure chaotic and
nor pure regular. 
Several theoretical NNSDs were suggested for interpretations of the 
experimental NNSDs. The most popular is, e.g., the
Brody distribution 
\cite{10,15}
\begin{equation}\label{7}
p(s) = A_{q\, }(1+q) s^{q} \exp (-A_{q} s^{q+1}), 
\end{equation}
where $q $ is an unique fitting parameter.  
The normalization constant is given by
\begin{equation}\label{8}
A_{q} =(1+q)\left[ {\Gamma \left( {\frac{q+2}{q+1}} \right)} \right]^{q+1}~,
\end{equation}
where $\Gamma(x)$ 
is the Gamma function. In the limit $q\to 0$, one has the 
Wigner distribution and for $q\to 1$, one finds the Poisson 
limit [see Eq.~(\ref{1})].

The new LWD approach is based on the expression
for 
NNSD $p(s)$ within the 
Wigner-Dyson theory \cite{21,22},
\begin{equation}
\label{9}
p(s)=A_{\rm{LWD}}^{-1}~g(s)\,\exp \left( {-\int\limits_0^s {d{s}'g({s}')} } \right),
\end{equation}
where $g(s)$ is the repulsion level density, 
linear in $s$,
\begin{equation}\label{10}
g(s)=a+bs,
\end{equation}
$a$ and $b $ are fitting parameters, and $A_{\rm{LWD}} $
is the normalization constant. It can be 
expressed analytically in terms of the error functions by using the 
normalization conditions. This is the linear Wigner-Dyson (LWD) 
two-parametric approach \cite{22}. 
Constants $a$ and $b$ can be
related by 
normalization conditions keeping, however, the quantitative measure of the 
separate Poisson and Wigner
contributions. 
Then, one gets
$A_{\rm{LWD}}=1$ and the 
one-parametric LWD distribution \cite{23} takes the form:
\begin{equation}
\label{11}
p(s)=[a(w)+b(w)s]\,\exp [-a(w)s\,-\,b(w)s^{2}/2]\quad ,
\end{equation}
where
\bea\l{12}
a(w)&=&\sqrt \pi \,w\,\exp 
(w^{2})\,\mbox{erfc(}w\mbox{)},\nonumber\\
b(w)&=&\frac{\pi }{2}\,\exp 
(2w^{2})\,\mbox{erfc}^{2}(w).
\eea
For the limit $w\to \infty$ ~($a\to 1$ and $b\to 0$), one obtains the Poisson 
distribution while for $w\to 0$ ($a\to 0$ and $b\to \pi /2)$, one 
arrives at the Wigner distribution [see Eq.~(\ref{1})].
\clearpage

\section{Discussions of the results}

\begin{figure*}
\includegraphics[width=18.0cm]{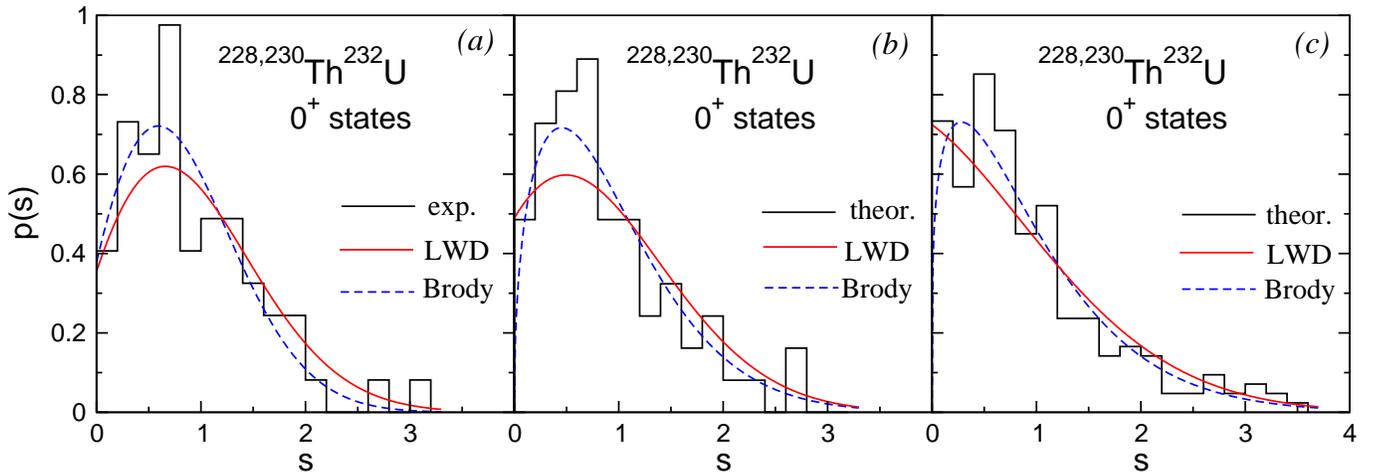}

\vspace{1.7cm}
\caption{{\small
    NNSDs for the experimental data (a) and the 
theoretical QPM results (b) in the same energy interval up to 3 MeV in  
$^{228,230}$Th and $^{232}$U actinide nuclei, and those (c) up to 4.2MeV. 
Other notations are the same as in Figs.~\ref{fig12} and \ref{fig13}.
}}
\label{fig14}
\end{figure*}

\begin{figure*}
  \includegraphics[width=18.0cm]{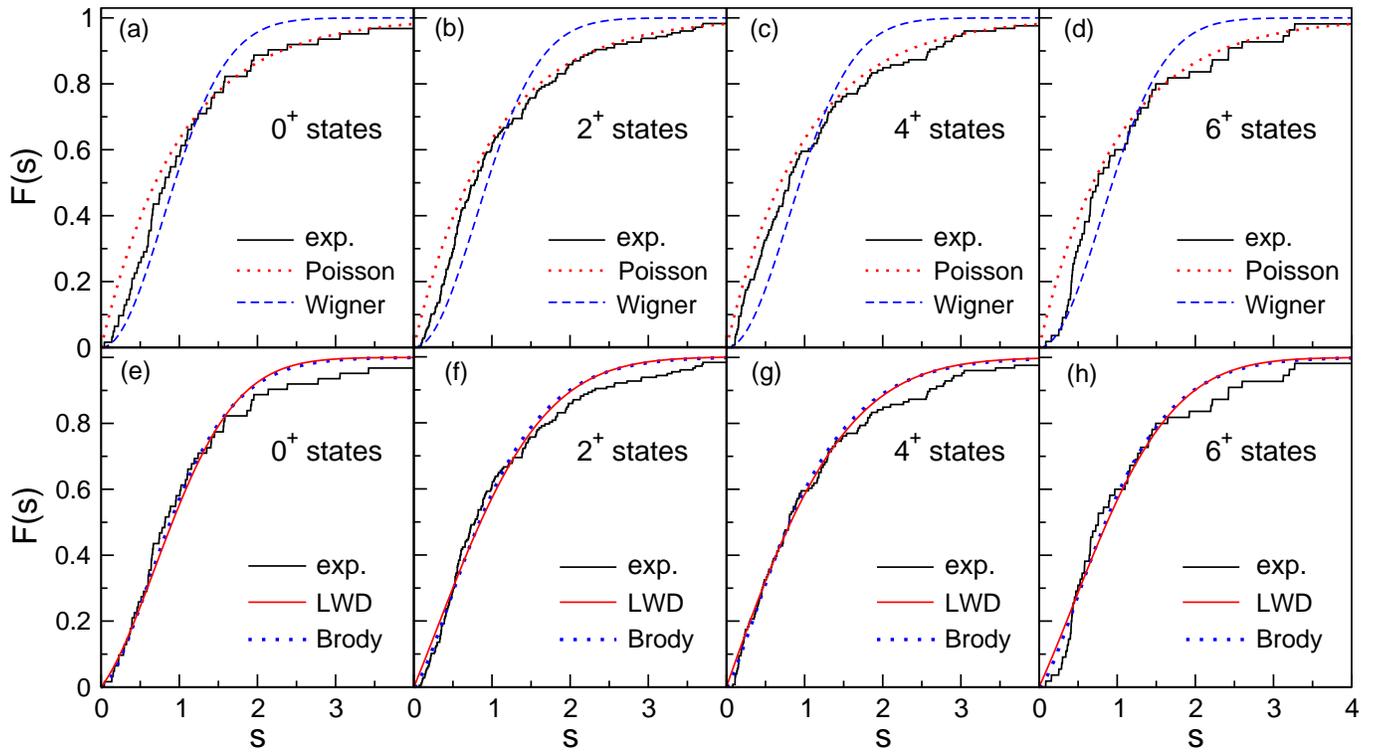}

\vspace{0.5cm}
\caption{{\small
    Cumulative distributions: upper line 
    (a)-(d) shows the 
comparison of the same experiment as in Fig.~\ref{fig14} 
with Poisson and Wigner 
limits (\ref{1}); 
the lower line 
(e)-(h) presents
the comparison with the LWD (\ref{11}) 
and Brody (\ref{7}) 
approach. 
}}
\label{fig15}
\end{figure*}

\begin{figure*}
\includegraphics[width=18.0cm]{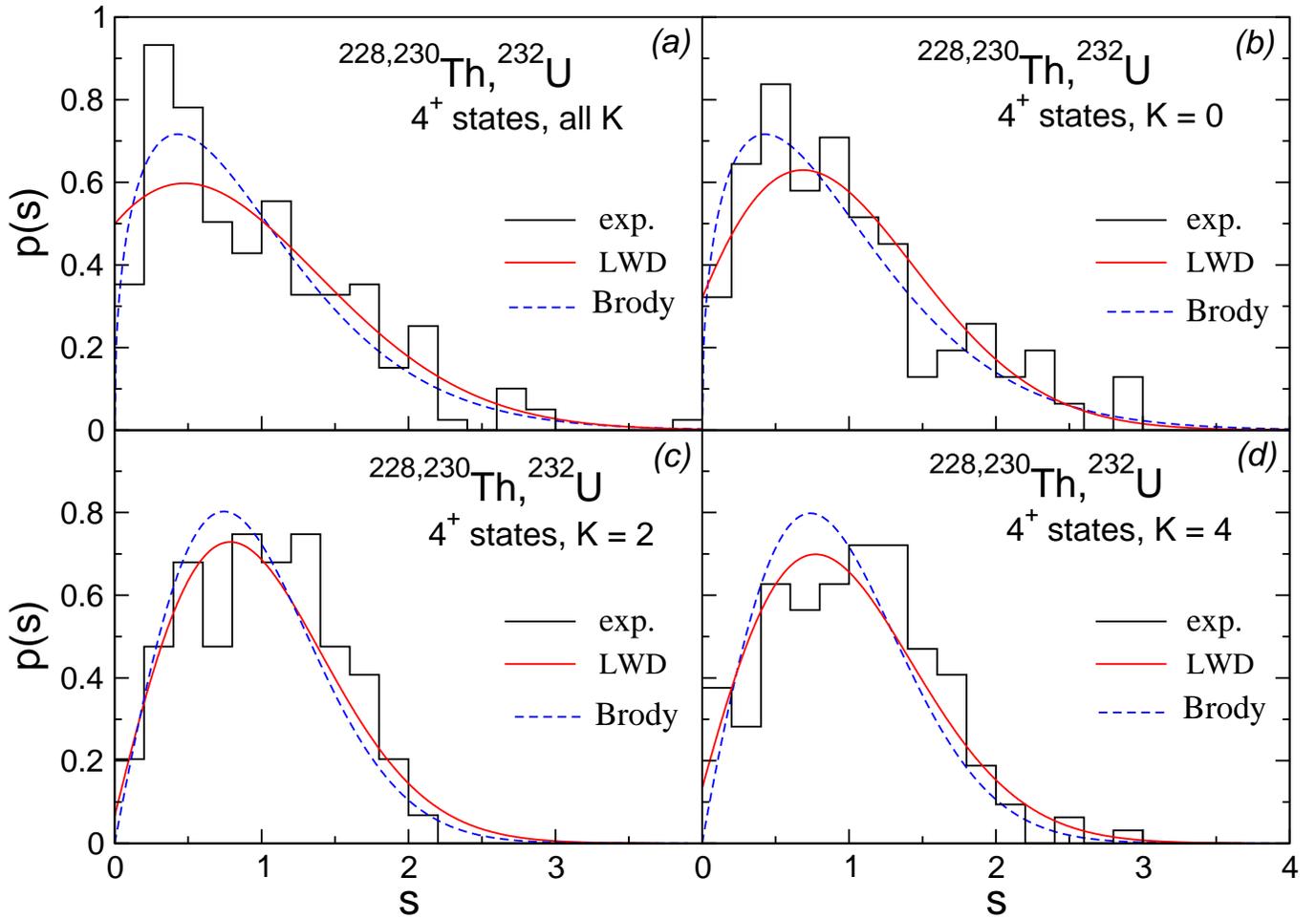}

\vspace{0.8cm}
\caption{{\small
 NNSDs for full spectrum (a) and symmetry breaking by fixed $K=0$ 
(b), 2 (c) and 4 (d) projections of the angular momentum 4$^{+}$ for the 
actinides which are included in Fig.~\ref{fig13}; 
red solid 
and blue dashed lines are fits by 
the LWD (\ref{11}) 
and Brody (\ref{7}) 
NNSDs, respectively.
}}
\label{fig16}
\end{figure*}

\vspace{0.5cm}
Fig.~\ref{fig4} shows good agreement of the one-parametric LWD approximation
  (\ref{11}) to the NNSD (\ref{9}) with the corresponding numerical (a,b)
  and experimental NDE (c)
distributions, also with their Poisson (a) and Wigner (b,c) limits (\ref{fig1}).
Other cases of the mixed order-chaos NNSDs
are presented in Figs.~\ref{fig12}-\ref{fig16}.
\vspace{6.0cm}

\vspace{13.0cm}
%

\vspace{-19.0cm}
Experimental NNSDs fitted by the LWD approximation for the 
collective states excited in several 
rare-earth nuclei (12 nuclei, 128 states for energies 
$E < 3$ MeV (a): $a =0.43$, $b =0.77$) and 
for $^{158}$Gd and $^{168}$Er (2 nuclei, 58 states for energies 
$E < 4.5$ MeV~(b): $a =0.82$, $b = 0.20$ {\bf )} are
shown in Fig.~\ref{fig12}. 
As seen from the comparison of (a) and 
(b) in Fig.~\ref{fig12}, 
one finds an intermediate chaos-order behavior between the 
Wigner and Poisson limits. A shift of these experimental and theoretical 
NNSDs from the Wigner to Poisson contributions is clearly shown in this 
figure from left to right, that is related to the increasing of lengths of 
the collective energy spectrum.

Fig.~\ref{fig13} shows the NNSDs for actinides, depending on the angular 
momentum $J=$0$^{+}$- 6$^{+}$
(4 nuclei, 438 states, namely 0$^{+}$ states: $a = 0.32$, $b = 0.98$; 
2$^{+}$ states: $a = 0.55$, $b = 0.59$; 4$^{+}$ states: $a = 0.67$, $b =
0.41$; 6$^{+}$ states: $a = 0.41$, $b = 0.81$). As seen from Fig.~\ref{fig13}, 
one finds a shift of the Wigner to Poisson contributions with increasing 
the angular momentum $J$ from 0$^{+}$ to  4$^{+}$.
Then, this shift slightly goes 
back to the Wigner limit because of missing levels \cite{43} due to very small 
cross-sections in the two-neutron transfer-reaction experiments at 6$^{+}$.


Fig.~\ref{fig14} 
shows the comparison of the experimental and theoretical (QPM) NNSDs 
for 
collective states 0$^{+\, }$ in a few actinide nuclei $^{228}$Th, 
$^{230}$Th and $^{232}$U. Experimental (a) and theoretical QPM (b) results 
are presented for energies $E <  3$MeV ($a = 0.36$, $b = 0.91 $ and 
$a = 0.49$, $b = 0.69$, relatively) 
and theoretical QPM (c) calculations
%
for energies in a wider region 
$E < 4.5$ MeV ($a= 0.72$, $b =0.33$). This figure presents 
completeness and collectivity 
of the used spectra because of good agreement of NNSDs between 
plots in (a) and (b). The comparison of (a,b) with (c) confirms the general 
law of a shift of NNSDs from the Wigner to the Poisson contribution with 
increasing the total energy interval.

Cumulative distributions,

\vspace{-0.5cm}
\begin{equation}
\label{13}
F(s)=\int\limits_0^s {d{s}^\prime p({s}^\prime} ),
\end{equation}

are shown in Fig.~\ref{fig15} 
for actinides (4 nuclei, 438 states) \cite{22}. 
As presented by 
this figure in (a-d), for all the angular momenta, the Wigner cumulative 
distribution well reproduces the behavior of empiric values at  
small spacing $s $ while the Poisson distribution is better 
fitted these data at larger 
$s$. A good comparison of these data with LWD (\ref{11}) and Brody
(\ref{7}) NNSDs
is shown in lower 
plots (e-h).

Fig.~\ref{fig16} shows the symmetry breaking phenomenon 
for the actinide nuclei with 
mixing all
projections $K$ of the angular momentum 4$^{+}$ (a) and fixing $K=0$ (b), 2 (c) 
and 4 (d). As the angular momentum projection $K$ is fixed in (b-d), one 
observes mainly a shift of the NNSDs to the chaotic Wigner contribution, in 
agreement with the results for the s.p. spectra \cite{21}. 
It is transparently shown 
in Sec. 2 through the comparison of 
Poincare sections in 
Figs.~\ref{fig1} and \ref{fig2}.

\section{Summary}

The statistical analysis of the spectra of collective states in the deformed 
rare and actinide nuclei has been presented. The experimental data obtained 
from the two-neutron transfer reactions are discussed.
The new 
method of the analysis of distributions of spacing intervals between the 
nearest neighbor levels (NNSDs) is suggested. This method has obvious 
advantages above the popular Brody method as giving the separate Wigner and 
Poisson contributions into the statistics of quantum spectra. Our LWD NNSDs 
can be also derived properly within the Wigner-Dyson theory, in contrast to 
the heuristic Brody approach. We found an intermediate behavior between the 
order and chaos in structures of quantum spectra of the collective states in 
terms of the Poisson and Wigner contributions. We observed a shift of  
NNSDs to the Poisson contributions with increasing the energy interval of 
these spectra and the angular momenta, that is in agreement with the 
random-matrix theory results. Statistical analysis of the cumulative 
distributions yields a relative role of the order and chaos depending on the 
spacing variable $s$. The symmetry-breaking effect with fixing the angular 
momentum projections $K$ of the shift of 
NNSDs to a more chaotic behavior 
(larger Wigner contribution) is a general property for the collective and 
single-particle states. 

This review might be helpful for understanding the order-chaos transitions 
in the collective spectra of strongly deformed nuclei. As perspectives, we 
are planning to study more properly and systematically these statistical 
properties of the nuclear collective states. The most attractive subject in 
these studies is the symmetry-breaking phenomena. 

\acknowledgments

We are grateful to K.~Arita, D.~Bucurescu,
S.~Mizutori, V.A.~Plujko, S.V.~Radionov, P.~Ring, A.I. Sanzhur
and M.~Spieker for many
helpful discussions.
This work was supported by the  budget program  ``Support for the development
of priority areas of scientific researches'' (Code 6541230).
One of us (A.G.M.) is 
also very grateful for kind hospitality during his working visits of 
Physical Department of the Nagoya Institute of Technology, also the Japanese 
Society of Promotion of Sciences for financial support, Grant No. S-14130.

\end{document}